\def\PRD{{\em Phys. Rev.} D}
\def\PRL{{\em Phys. Rev. Letts.} }
\def\NIMA{{\em Nucl. Ins. Meth.} A}
\documentstyle[epsf,aps,prd,besnote]{revtex}
\input{psfig}
\begin{document}
\title {Measurement of the Inclusive Charm Cross Section at 4.03 GeV and 4.14 GeV}
% Following can be inserted into revtex document.

\author{
J.~Z.~Bai,$^1$   Y.~Ban,$^5$      J.~G.~Bian,$^1$
I.~Blum,$^{12}$ 
G.~P.~Chen,$^1$  H.~F.~Chen,$^{11}$  
J.~Chen,$^3$ 
J.~C.~Chen,$^1$  Y.~Chen,$^1$ Y.~B.~Chen,$^1$
% ...
\cite{atNU0}
Y.~Y.~Wang,$^1$  
M.~Weaver,$^2$
C.~L.~Wei,$^1$   N.~Wu,$^1$       Y.~G.~Wu,$^1$
D.~M.~Xi,$^1$    X.~M.~Xia,$^1$   P.~P.~Xie,$^1$    Y.~Xie,$^1$
Y.~H.~Xie,$^1$   G.~F.~Xu,$^1$    S.~T.~Xue,$^1$
J.~Yan,$^1$      W.~G.~Yan,$^1$   C.~M.~Yang,$^1$   C.~Y.~Yang,$^1$
H.~X.~Yang,$^1$  J.~Yang,$^1$     
W.~Yang,$^3$
X.~F.~Yang,$^1$  M.~H.~Ye,$^1$    S.~W.~Ye,$^{11}$
Y.~X.~Ye,$^{11}$   C.~S.~Yu,$^1$    C.~X.~Yu,$^1$     G.~W.~Yu,$^1$
Y.~H.~Yu,$^4$    Z.~Q.~Yu,$^1$    C.~Z.~Yuan,$^1$   Y.~Yuan,$^1$
B.~Y.~Zhang,$^1$  C.~Zhang,$^1$   C.~C.~Zhang,$^1$  D.~H.~Zhang,$^1$  
Dehong~Zhang,$^1$
H.~L.~Zhang,$^1$ J.~Zhang,$^1$    J.~W.~Zhang,$^1$  L.~Zhang,$^1$
L.~S.~Zhang,$^1$ P.~Zhang,$^1$
Q.~J.~Zhang,$^1$ S.~Q.~Zhang,$^1$ X.~Y.~Zhang,$^6$  Y.~Y.~Zhang,$^1$
D.~X.~Zhao,$^1$  H.~W.~Zhao,$^1$  Jiawei~Zhao,$^{11}$ J.~W.~Zhao,$^1$
M.~Zhao,$^1$     W.~R.~Zhao,$^1$  Z.~G.~Zhao,$^1$   J.~P.~Zheng,$^1$
L.~S.~Zheng,$^1$ Z.~P.~Zheng,$^1$ B.~Q.~Zhou,$^1$   G.~P.~Zhou,$^1$
H.~S.~Zhou,$^1$  L.~Zhou,$^1$     K.~J.~Zhu,$^1$    Q.~M.~Zhu,$^1$
Y.~C.~Zhu,$^1$   Y.~S.~Zhu,$^1$   B.~A.~Zhuang$^1$
\\ (BES Collaboration)}

\address{
$^1$Institute of High Energy Physics, Beijing 100039, People's Republic of
 China\\
$^2$California Institute of Technology, Pasadena, California 91125\\
$^3$Colorado State University, Fort Collins, Colorado 80523\\
$^4$Hangzhou University, Hangzhou 310028, People's Republic of China\\
$^5$Peking University, Beijing 100871, People's Republic of China\\
$^6$Shandong University, Jinan 250100, People's Republic of China\\
$^7$Shanghai Jiaotong University, Shanghai 200030, People's Republic of China\\
$^8$Stanford Linear Accelerator Center, Stanford, California 94309\\
$^9$University of Hawaii, Honolulu, Hawaii 96822\\
$^{10}$University of California at Irvine, Irvine, California 92717\\
$^{11}$University of Science and Technology of China, Hefei 230026,
People's Republic of China\\
$^{12}$University of Texas at Dallas, Richardson, Texas 75083-0688}

\date{\today}
\twocolumn[
% \begin{abstract}
The cross section for charmed meson production at $\sqrt{s} = 4.03 $ and
$4.14 $ GeV has been measured with the Beijing Spectrometer.  
The measurement 
was made using 22.3 $pb^{-1}$ of $e^+e^-$ data collected at 
4.03 GeV and 1.5 $pb^{-1}$ of $e^+e^-$ data collected at 
4.14 GeV.  Inclusive observed cross sections for the production of 
charged and neutral $D$ mesons
and momentum spectra
are presented.  Observed cross sections were radiatively
corrected to obtain tree level cross sections.  Measurements of 
the total hadronic cross section are obtained from the charmed meson cross section and an extrapolation of results from below the charm 
threshold.
% \end{abstract}

\maketitle
]
\section{Introduction}
The hadronic cross section of $e^+ e^-$ at all energies is needed to calculate
the effects of vacuum polarization on parameters of the Standard Model.
The energy region which contributes the largest uncertainty is the charm 
threshold region
where the hadronic cross section has only been measured with an accuracy of 
$15-20\%$\cite{swartz}.  
Traditionally, $\sigma_{hadron}$ is measured by counting hadronic events.  
This method requires a detailed 
understanding of trigger conditions, the efficiency of hadronic event
selection criteria, and a subtraction of two photon events and other backgrounds.  An alternative is measuring $\sigma_{charm}$ 
and adding this to an extrapolation of the 
$\sigma_{u,d,s}$ contribution from the region below charm threshold.  The 
charmed mesons used in this study are the $D^0$ and $D^+$.  The $D_s$ cross sections are taken from earlier works.  There is no evidence for continuum charmonium production\cite{charm}.
The charm counting method
is intrinsically less sensitive to trigger conditions, beam-related
backgrounds, and two photon backgrounds due to the distinctive 
topology of charmed meson events.

\section{Data Selection}
The data used for this analysis were accumulated 
with the Beijing Spectrometer\cite{bes:nucl}; the total integrated 
luminosity at 4.03 (4.14) GeV was 22.3 (1.5) $pb^{-1}$. Candidate tracks 
were required to have a good track fit passing within 1.5 cm of the 
collision point in $R$ and 15 cm in $z$, and satisfying $|\cos\theta|<0.85$.
Particle identification was provided by 
an array of time of flight scintillation counters (TOF) and 
specific ionization measurements in the drift chamber used for 
charged particle tracking ($dE/dx$).  For pions, consistency 
$(CL(\pi)>0.1\%)$ and loose electron rejection $(L_\pi/(L_\pi+L_e)>0.2)$ were
required, where $L_X$ is the likelihood for hypothesis $X$.  Kaon identification 
required consistency $(CL(K)>0.1\%)$, pion rejection $(L_K/(L_K+L_\pi)>0.5)$ 
and loose electron rejection $(L_\pi/(L_\pi+L_e)>0.2)$.  Multiple 
counting of $D^0$ candidates was removed by positively identifying pions 
using the selection $(L_K/(L_K+L_\pi)<0.5)$.  Muons are rejected using a 
momentum dependent criteria based on track penetration into the muon detector.

\section{$D^0$ and $D^+$ Signal}
Inclusive $K^- \pi^+$ and $K^- \pi^+ \pi^+$ invariant mass distributions are shown in
Figs.~\ref{fig:1} and~\ref{fig:2} for 4.03 and 4.14 GeV, 
respectively.  (Here and throughout this paper, reference to a state also 
implies its charge conjugate state.)
Each histogram was fit to a function which included a Gaussian signal plus
a background function.  For the $K \pi$ distributions (Figs.~\ref{fig:1}a and~\ref{fig:2}a), the background function 
consisted of a Gaussian centered at 1.60 GeV to account for the contribution to the $K \pi$ spectrum from $D^0 \rightarrow 
K \pi \pi^0$ decays plus a third order
polynomial background.  For the $K \pi \pi$ distributions (Figs.~\ref{fig:1}b and~\ref{fig:2}b), the background
function consisted of a third order polynomial.  The number of signal
events in each mode is given in Table~\ref{sig:1}.

The momentum spectra of $D$ candidates with a mass between 1.81 and 1.91 GeV 
is presented in Fig.~\ref{fig:3} for $\sqrt{s} = 4.03$ GeV.  The three momentum 
regions near 0.15, 0.55, and 0.75 GeV correspond to $D^*\overline{D}^*$,
$D^*\overline{D}$, and $D\overline{D}$ production, respectively.
The spectrum for $\sqrt{s} = 4.14$ GeV is shown in Fig.~\ref{fig:4}.
The momentum regions near 0.45 and 0.70 and 0.90 GeV correspond
to $D^*\overline{D}^*$, $D^*\overline{D}$ and $D\overline{D}$ production, respectively.  The shapes of the $D^*\overline{D}^*$ spectrum and part of the $D^*\overline{D}$ spectrum
are broadened due to Doppler-smeared $D$ mesons coming from $D^*$ decays.
In addition, a small low momentum tail on each structure is expected due to
initial state radiation.  The background shape under the momentum spectrum
is not flat, making a direct subtraction difficult.

\section{Cross Section}
If the reconstruction efficiency for $D$ mesons were constant with respect to 
momentum, the observed cross section could be determined using \begin{eqnarray}
\sigma(e^+e^- \rightarrow D X) = \frac{N(\rm signal\em )}
{\epsilon B \cal L\em},
\end{eqnarray}
where $N$(signal) is the number of signal events, $\epsilon$ is the
efficiency, $B$ is the branching fraction of the $D$ meson to a decay
mode and $\cal L\em$ is the luminosity.  However, Monte Carlo studies show some momentum dependence 
to the reconstruction efficiency (Fig.~\ref{fig:5}).  In order to measure the 
cross section, the momentum spectrum for $D^0$ and $D^+$ mesons from 50 to 850 MeV was 
divided into 20 (40) MeV slices for 4.03 (4.14) GeV data.  For each momentum slice, the invariant mass 
distribution was fit with a Gaussian plus a polynomial background.  
The central value of the Gaussian was fixed at the nominal $D$ mass; the width was fixed to a 
momentum dependent value determined by a fit to a coarser slicing of the data 
(Fig.~\ref{fig:6}).  The differential cross section with respect to momentum 
for this data is shown in Figs.~\ref{fig:7} and~\ref{fig:8} for 4.03 and 
4.14 GeV, respectively.  The cross section times branching fractions of 
$D^0$ and $D^+$ mesons and cross sections calculated using the branching 
fractions of ref.\cite{pdg} are shown in Table~\ref{sigma:1}.
The $\sigma \cdot B$ values from this measurement are compatible with 
previous measurements by 
Mark I\cite{markIsigB77}\cite{markIsigB79}
and 
Mark II\cite{coles}
as shown in Fig.~\ref{prev:1}.

\section{Corrections for Initial State Radiation}
The tree level cross sections for charm at $\sqrt{s} = 4.03$ GeV and 
$\sqrt{s} = 4.14$ GeV were obtained by correcting the observed cross 
section for the effects of initial state radiation (ISR). The ISR correction 
is dependent on the cross section for 
all energies less than the nominal energy.  Since these measurements were 
performed only at two energies, some theoretical modeling of the cross 
section distribution was required.  Two different theoretical predictions for these cross sections were
used in this analysis, the Coupled Channel Model\cite{eic:phys}, and a
$P$-wave phase space formalism.
The models provide predictions for $\sigma_{B,i}(s_{\rm eff})$, the tree level
(Born) cross section as a
function of the effective center of mass energy squared for production mode
$i$, where $i = D^0\overline{D}^0$, $D^+D^-$, $D^{*0}\overline{D}^0$, 
$D^{*+}D^-$, $D^{*0}\overline{D}^{*0}$, $D^{*+}D^{*-}$.  The
effective center of mass energy squared is given by
\begin{equation}
s_{\rm eff} \equiv s_{\rm nom}(1-k),
\end{equation}
where $k \cdot E_{\rm beam}$ is the energy of radiated photons and $s_{\rm nom}$
is the nominal center of mass energy squared.  The tree level cross sections 
are convoluted with a sampling function $f(k, s_{\rm nom})$ that represents 
a first order calculation of the effective luminosity for radiated photons in a two
body radiation model\cite{isr}, giving the observed cross section at $s_{\rm nom}$:

\begin{eqnarray}
\lefteqn{\sigma_{\rm obs,i}(s_{\rm nom}) =}  \nonumber \\
& \int_0^1 dk \cdot f(k,s_{\rm nom \em}) 
\sigma_{B,i} (s_{\rm eff})(1+\delta_{\rm VP}(s_{\rm eff})).
\end{eqnarray}
The vacuum polarization correction $(1+\delta_{\rm VP})$ includes both
leptonic and hadronic terms. It varies from charm threshold to 4.14 GeV 
by less than $\pm$2\%.  
It is treated as a constant with the value

\begin{equation}
(1+\delta_{\rm VP})=1.047 \pm 0.024
\end{equation}
and moved outside the integrand. 
With this simplification, the ISR correction, 
$g_i(s_{\rm nom})$, is defined by
\begin{equation}
g_{i}(s_{\rm nom}) \equiv
\frac{\sigma_{{\rm obs,} i }(s_{\rm nom})}
{\sigma_{B,i}(s_{\rm nom})(1+\delta_{\rm VP})}.
\end{equation}

The $D^{*0}$ and $D^{*+}$ branching fractions are used to calculate $N_{D^0,i}$ and
$N_{D^+,i}$, the mean number of $D^0$ and $D^+$ per production mode $i$ event.  The
$N_{D^0,i}$ and $N_{D^+,i}$ values are used to weight the ISR correction for
each
mode to obtain the ISR correction averaged over all production modes:

\begin{equation}
g_{D^{0}}(s_{\rm nom})  = \frac{\sum_{i} g_i(s_{\rm nom}) N_{D^0,i}}
{\sum_{i} N_{D^0,i}}
\end{equation}

\begin{equation}
g_{D^{+}}(s_{\rm nom})  = \frac{\sum_{i} g_i(s_{\rm nom}) N_{D^+,i}}
{\sum_{i} N_{D^+,i}}.
\end{equation}

This procedure results in the corrections shown in Fig.~\ref{isr:5}.  
Since neither method models the data precisely, and the two 
models vary differently with energy, a systematic uncertainty is assigned
to be one half the rms difference between the two models over the 
energy range 3.9 GeV to 4.2 GeV excluding the region from 4.021 to 4.027
where the Coupled Channel Model $D\overline{D}$ cross section is tiny, causing
the ISR correction to diverge.  The corrections for initial 
state radiation at $\sqrt{s}=4.03$ GeV and $\sqrt{s}= 4.14$ GeV are:
\begin{eqnarray}
 4.03\; {\rm GeV}\!\!:& \; g_{D^0} = 0.67 \pm 0.05 \\ & 
g_{D^+} = 0.73 \pm 0.05 \\
4.14\; {\rm GeV}\!\!:& \; g_{D^0} = 0.83 \pm 0.06 \\ & 
g_{D^+} = 0.84 \pm 0.06.  
\end{eqnarray} 

The ISR correction for $D_s$ mesons was calculated using the Coupled Channel 
Model and the same $P$-wave phase space formalism.  
Figure~\ref{isr:ds} shows a prediction for the tree level cross section of 
$D_s$ and $D_s^*$, the observed cross 
section of $D_s$ and $D_s^*$, the fraction of $D_s$ mesons from direct 
production, and the fraction of $D_s$ mesons from $D_s^*$ decays.  
Figure~\ref{isr:5}c shows the ISR correction 
for $D_s$ mesons.  The ISR contribution to the systematic error 
for $D_s$ production is taken as one half the rms difference between the two models:
\begin{eqnarray}
4.03\; {\rm GeV}\!\!:&\; g_{D_s} = 0.73 \pm 0.04 \\
4.14\; {\rm GeV}\!\!:&\; g_{D_s} = 0.78 \pm 0.05. 
\end{eqnarray}

The ISR and vacuum polarization corrections are applied to the 
observed $D^0$ and $D^+$ cross sections found in Table~\ref{sigma:1}
to obtain the tree level cross section for $D^0$ and 
$D^+$ at 4.03 GeV and 4.14 GeV as shown below:
\begin{eqnarray}
4.03\; {\rm GeV}\!\!:&\; \sigma_{D^0} + \sigma_{\overline{D}^0} = 19.9 \pm 0.6 \pm 2.3 \, {\rm nb} \\
      & \sigma_{D^+} + \sigma_{D^-} = 6.5 \pm 0.2 \pm 0.8 \, {\rm nb}  \\
4.14\; {\rm GeV}\!\!:&\; \sigma_{D^0} + \sigma_{\overline{D}^0} =  9.3 \pm 2.1 \pm 1.1 \, {\rm nb} \\
      & \sigma_{D^+} + \sigma_{D^-} = 1.9 \pm 0.9 \pm 0.2 \, {\rm nb}.   
\end{eqnarray}
Systematic uncertainties are treated in Section VI.

The BES observed $D_s$ cross section at 4.03 GeV is 
$\sigma_{D_s^+} + \sigma_{D_s^-} = 0.62 \pm 0.12 \pm 0.20$ nb\cite{bes:phys}.  
After applying the ISR and vacuum polarization corrections, the tree level $D_s$ cross section is 
$\sigma_{D_s^+} + \sigma_{D_s^-} = 0.81 \pm 0.16 \pm 0.27$ nb.
The Mark III observations of $D_s$ at 4.14 GeV give 
$\sigma_{D_s^+} + \sigma_{D_s^-} = 1.34 \pm 0.32 \pm 0.34$ nb\cite{mark:phys}.  
After correction, the tree level value is 
$\sigma_{D_s^+} + \sigma_{D_s^-} = 1.64 \pm 0.39 \pm 0.42$ nb.

\section{Systematics}
Several systematic checks were performed.  The numbers of signal events in
the distributions shown in Figs.~\ref{fig:1}, and~\ref{fig:2} were compared to the sum of 
signal events from each momentum slice as shown in Table~\ref{tab:1}.
Good agreement for the 4.03 GeV data validates the 
slicing technique.  For the 4.14 GeV data set, which is much smaller, the agreement is poorer.  This 
could be due to statistical fluctuations.

The analysis was repeated using wrong sign combinations $(K^+\pi^+, K^+
\pi^+\pi^-)$ to explore systematic bias from the slicing and fitting procedure.
As expected, the structures that are so evident in the right sign spectra
are absent in the wrong sign spectra. (Fig.~\ref{fig:9}).  There is, 
however, a small excess when the $K^- \pi^-\pi^+$ spectrum is integrated: 
\begin{eqnarray}
(\sigma_{D^0} + \sigma_{\overline{D}^0})_{WS} = 0.29 \pm 0.21 \, {\rm nb} \\ 
 (\sigma_{D^+} + \sigma_{D^-})_{WS} = 0.7 \pm 0.2 \, {\rm nb}. 
\end{eqnarray}
Both the data and the charged MC show a small excess that was not present in 
neutral decays.  A much larger MC study would be required to prove whether 
the source is procedural, or if there
is a small feed down from right sign channels into the wrong sign analysis.
If the source of the excess is procedural, the MC efficiency calculation
should correct for the effect in the data.  In either case, no correction
is required.

Systematic errors arising from the choice of parameters were evaluated by repeating the analyses using
different bin sizes, fitting ranges, and background function shapes. 
Electron particle identification dominates the integrated luminosity uncertainty as determined from wide angle Bhabha scattering events.  The uncertainty is evaluated by comparing samples selected independently using $dE/dx$ and the barrel shower counter.
In addition there are systematic errors
due to the uncertainties in the Monte Carlo-determined reconstruction efficiency, 
errors in the charmed meson branching fractions, and the uncertainties in the evaluation 
of the ISR correction scheme as discussed above.  Magnitudes of these 
systematic uncertainties are shown in Table~\ref{par:1}.

Sources of systematic uncertainty are segregated into components that are common 
or independent for $D^0$, $D^+$, and $D_S$ measurements.
The common components are the
integrated luminosity measurement, the 
ISR correction, the vacuum polarization correction, 
and a portion of the $D$ branching fraction uncertainties.  
Since the absolute branching fraction scale for $D^+$ mesons depends
on the $D^0$ branching fraction scale,
the total percentage uncertainty for $D^+$ branching fraction (6.7\%) 
is split into a common component 
that matches the percentage uncertainty for the $D^0$ branching fraction 
(2.3\%) and an independent
component (6.2\%).  All other systematic uncertainties are treated as 
independent and added in quadrature. 
Values are found in Table~\ref{par:1}.  

The total observed $D^0$ and $D^+$ cross sections are shown in
Table~\ref{sigma:1}.  Tree level $D^0$ and $D^+$ cross sections are shown
in Table~\ref{sigma:tree}.

\section{Total Inclusive Charm Cross Section}
Since all $D$ mesons are produced in pairs,
the tree level non-strange $D$ cross sections are:
\begin{eqnarray}
4.03 \; {\rm GeV}\!\!:& \;\sigma(D\overline{D}X)  
    & = 13.2 \pm 0.3 \pm 1.4 \, {\rm nb} \\
4.14 \; {\rm GeV}\!\!:& \;\sigma(D\overline{D}X) 
    & = 5.6 \pm 1.1 \pm 0.6 \, {\rm nb}. 
\end{eqnarray}
Adding the tree level $D$ cross sections to the tree 
level $D_s$ cross 
sections gives the total tree level charm cross section:
\begin{eqnarray}
4.03 \;{\rm GeV}\!\!: & \;\sigma_{\rm charm} &= 13.6 \pm 0.3 \pm 1.5 \, {\rm 
nb} \\
4.14 \;{\rm GeV}\!\!: & \;\sigma_{\rm charm} &= 6.4 \pm 1.2 \pm 0.7 \, {\rm nb}.
\end{eqnarray}
These results are compared to Coupled Channel model predictions in 
Table~\ref{sigma:tree}.

\section{Measurement of $R_D$ and $R$}
A measurement of $R_{D}$ is obtained by dividing $2\times\sigma_{\rm charm}$ by the 
QED prediction for the tree level muon pair cross section
\begin{equation}
\sigma(e^+e^-\rightarrow \mu^+\mu^-) = \frac{86.8 \, \rm nb}
                                         {\em s \, {\rm (GeV)^2}}
\end{equation}
giving:
\begin{eqnarray}
4.03\;{\rm GeV}\!\!: &\; R_{D} = 5.10 \pm 0.12 \pm 0.55 \\
4.14\;{\rm GeV}\!\!: &\; R_{D} = 2.53 \pm 0.46 \pm 0.27.
\end{eqnarray}
The contribution to $R$ in the charm threshold region 
from the light quarks, $R_{uds}$
is estimated to be $2.5 \pm 0.25$\cite{coles}. 
This value was compiled from an average of  measurements of $R$ 
below charm threshold\cite{plutoR}\cite{brand}\cite{markIR}.
The theoretical expectation is that $R_{uds}$ 
is approximately independent of center of mass energy in this region.
The value of $R$ is evaluated using $R = R_D/2 + R_{uds}$  
giving:
\begin{eqnarray}
4.03\; {\rm GeV}\!\!: &\; R = 5.05 \pm 0.06 \pm 0.37 \\
4.14\; {\rm GeV}\!\!: &\; R = 3.76 \pm 0.23 \pm 0.28.
\end{eqnarray}
This measurement is more precise, but compatible with previous 
$R$ measurements\cite{plutoR}\cite{markIR}
using the total cross section method shown in
Figures~\ref{prev:2}a and \ref{prev:2}b
and a previous measurement\cite{coles} employing 
a similar $R_D$ technique shown in Figure~\ref{prev:2}c.
Charm-counting complements direct $R$ measurements
since the two methods feature different systematic uncertainties.

We would like to thank the staffs of the BEPC accelerator and the Computing 
Center at the Institute of High Energy Physics (Beijing).  This work was 
supported in part by the National Natural Science Foundation of China under 
contract No.  19290400 and the Chinese Academy of Sciences under contract No.
KJ85 (IHEP); by the Department of Energy under contract Nos. DE-FG03-92ER40701
(Caltech), DE-FG03-93ER40788 (Colorado State University), DE-AC02-76ER03069 
(MIT), DE-AC03-76SF00515 (SLAC), DE-FG03-91ER40679 (UC Irvine), DE-FG03-94ER40833
(University of Hawaii), DE-FG03-95ER40925 (UT Dallas); by the U.S. National
Science Foundation Grant No. PHY9203212 (University of Washington).

\onecolumn
\begin{bestable}
\begin{tabular}{lcc}
Mode & $E_{c.m.}$(GeV) & Events \\
\hline
$D^0 \rightarrow K^-\pi^+$ & 4.03 & $4174 \pm 163$ \\
$D^+ \rightarrow K^-\pi^+\pi^+$ & 4.03 & $2341 \pm 138$ \\
$D^0 \rightarrow K^- \pi^+$ & 4.14 & $249 \pm 36$ \\
$D^+ \rightarrow K^-\pi^+\pi^+$ & 4.14 & $126 \pm 31$ \\
\end{tabular}
\caption{Number of inclusive signal events for each mode.}
\label{sig:1}
\end{bestable}
\begin{bestable}
\begin{tabular}{lcc}
& $\sqrt{s} = 4.03 $ GeV & $\sqrt{s} = 4.14 $ GeV \\
\hline
$(\sigma_{D^0} + \sigma_{\overline{D}^0}) 
\cdot B(D^0 \rightarrow K^- \pi^+) $ &
$ 0.537 \pm 0.015 \pm 0.047 $ nb & $ 0.31 \pm 0.07 \pm 0.03 $ nb \\
$(\sigma_{D^+} + \sigma_{D^-}) 
 \cdot B(D^+ \rightarrow K^- \pi^+ \pi^+)$ &
$ 0.449 \pm 0.017 \pm 0.036 $ nb & $ 0.15 \pm 0.07 \pm 0.01 $ nb \\
$\sigma_{D^0} + \sigma_{\overline{D}^0} $ & $ 13.9 \pm 0.4 \pm 1.3 $  nb & $ 8.1 \pm 1.8 \pm 0.7 $ nb \\
$\sigma_{D^+} + \sigma_{D^-} $ & $  5.0 \pm 0.2 \pm 0.5 $  nb & $ 1.7 \pm 0.8 \pm 0.2 $ nb \\
\end{tabular}
\caption{The observed cross section times branching fraction of $D^0$ and $D^+$ and the
observed cross section of $D^0$ and $D^+$ at 4.03 GeV and 4.14 GeV.}
\label{sigma:1}
\end{bestable}
\begin{bestable}
\begin{tabular}{ccccc}
&\multicolumn{2}{c}{$\sqrt{s} = 4.03 $ GeV} & \multicolumn{2}{c}{$\sqrt{s} = 4.14$ GeV} \\
& Inclusive $D$ & Sum of Slices 
& Inclusive $D$ & Sum of Slices \\ \hline
$D^0$ & $4174 \pm 163$ & $4232 \pm 114$ 
      & $ 249 \pm  36$ & $ 210 \pm  46$ \\ \hline
$D^+$ & $2341 \pm 138$ & $2395 \pm  94$ 
      & $ 126 \pm  31$ & $  66 \pm  33$ \\ \hline
\end{tabular}
\caption{Comparison of the number of events found from the inclusive mass
fit and the sum of the events from invariant mass fits for the momentum slices.}
\label{tab:1}
\end{bestable}
\begin{bestable}
\begin{tabular}{ccccccc}
Common & 4.03 $D^0$ & 4.03 $D^+$ & 4.14 $D^0$ & 4.14 $D^+$ & 4.03 $D_s$ & 4.14 $D_s$ \\
\hline
Luminosity 	         &5\%  &5\%  &5\%  &5\%  &5\%  &      \\
$D$ Branching Fraction	 &2.3\%&2.3\%&2.3\%&2.3\%&     &      \\ 
ISR 	                 &7\%  &7\%  &7\%  &7\%  &6\%  &6\%   \\ 
Vacuum Polarization      &2.2\%&2.2\%&2.2\%&2.2\%&2.2\%&2.2\% \\ 
\hline
\hline
Independent & 4.03 $D^0$ & 4.03 $D^+$ & 4.14 $D^0$ & 4.14 $D^+$ & 4.03 $D_s$ & 4.14 $D_s$ \\
\hline
$D$ Branching Fraction 	 &   &6.2\%&    &6.2\%&       &         \\
MC statistics 	         &5\%&4\%  &5\% &4\%  &       &         \\
Fit parameters	         &5\%&5\%  &5\% &5\%  &       &         \\
Previous measurements    &   &     &    &     &31.9\% &15.0\%   \\
\end{tabular}
\caption{Common and independent sources of systematic uncertainty.  
When summing cross sections for different $D$ species, 
common uncertainties are added linearly and 
independent uncertainties are added quadratically.}
\label{par:1}
\end{bestable}
\begin{bestable}
\begin{tabular}{lcc}
$\sqrt{s} = 4.03 $ GeV & Experiment & Coupled Channel Model \\
\hline
$\sigma_{D^0} + \sigma_{\overline{D}^0} $
& $ 19.9 \pm 0.6 \pm 2.3 $ nb 
& $ 18.2 $ nb \\
$\sigma_{D^+} + \sigma_{D^-} $ 
& $  6.5 \pm 0.2 \pm 0.8 $ nb 
& $  6.0 $ nb \\
$\sigma_{D_s^+} + \sigma_{D_s^-} $
& $ 0.81 \pm 0.16 \pm 0.27 $ nb
& $ 1.61 $ nb \\
$\sigma_{\rm charm} $ 
& $ 13.6 \pm 0.3 \pm 1.5 $ nb
& $ 12.9 $ nb \\
\hline
\hline
$\sqrt{s} = 4.14 $ GeV & Experiment & Coupled Channel Model \\
\hline
$\sigma_{D^0} + \sigma_{\overline{D}^0} $ 
& $  9.3 \pm 2.1 \pm 1.1 $ nb 
& $ 15.1 $ nb \\
$\sigma_{D^+} + \sigma_{D^-} $ 
& $  1.9 \pm 0.9 \pm 0.2 $ nb 
& $  4.5 $ nb \\
$\sigma_{D_s^+} + \sigma_{D_s^-} $
& $ 1.64 \pm 0.39 \pm 0.42 $ nb
& $ 1.85 $ nb \\
$\sigma_{\rm charm} $
& $  6.4 \pm 1.2 \pm 0.7 $ nb
& $ 10.7 $ nb \\
\end{tabular}
\caption{Comparison of tree level cross section measurements 
with predictions of the Coupled Channel Model.}
\label{sigma:tree}
\end{bestable}

\twocolumn
\begin{besfig}
\besfigscale{0.1}
\psfig{file=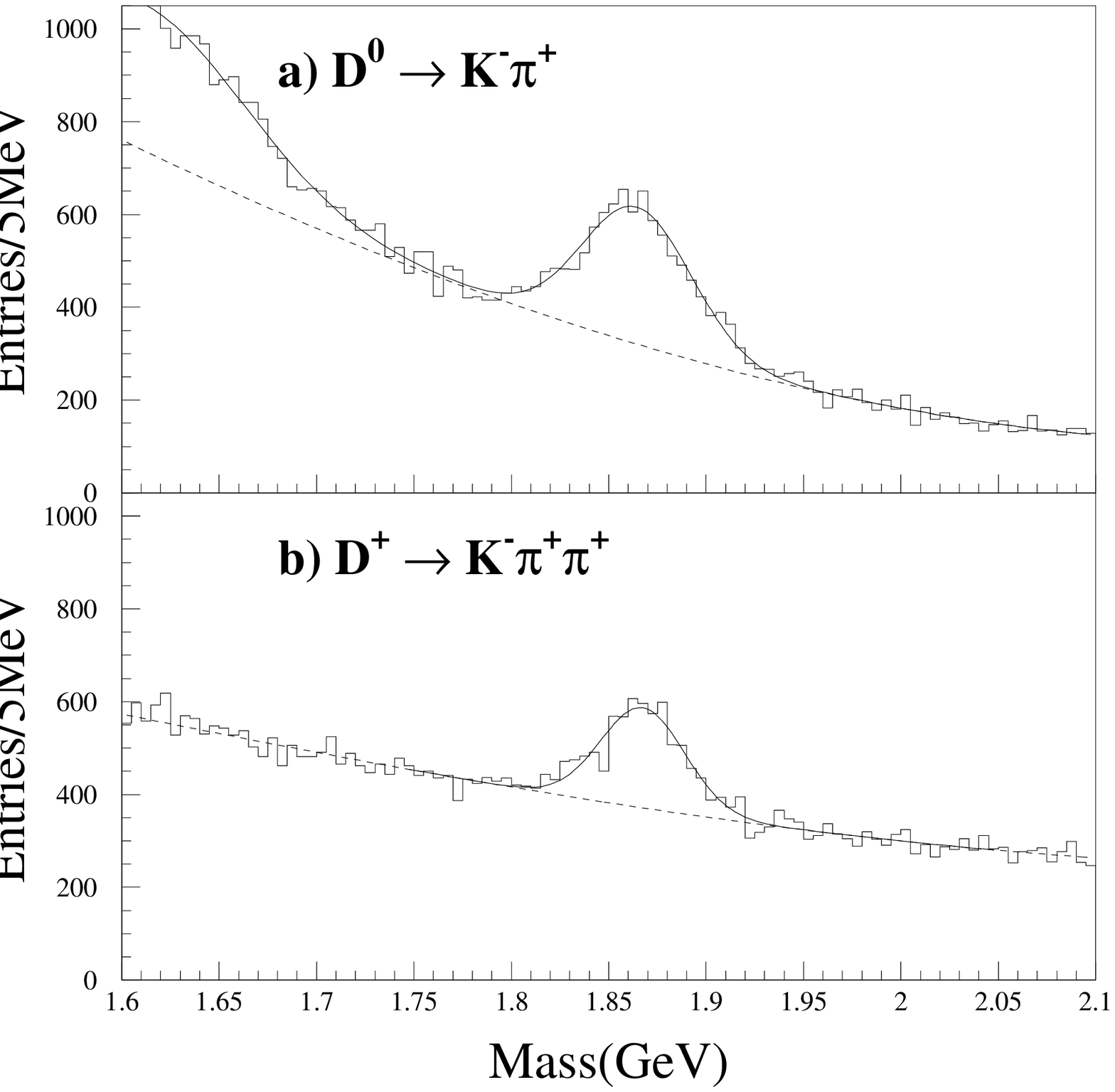,width=\textwidth,height=\textwidth}
\caption{The invariant mass of a) $K \pi$ and b) $K \pi \pi$ tags with 
a momentum between 0 and 1 GeV at $
E_{\rm c.m.}
 = 4.03$ GeV.} 
\label{fig:1}
\end{besfig} 
\begin{besfig}
\besfigscale{0.1}
\psfig{file=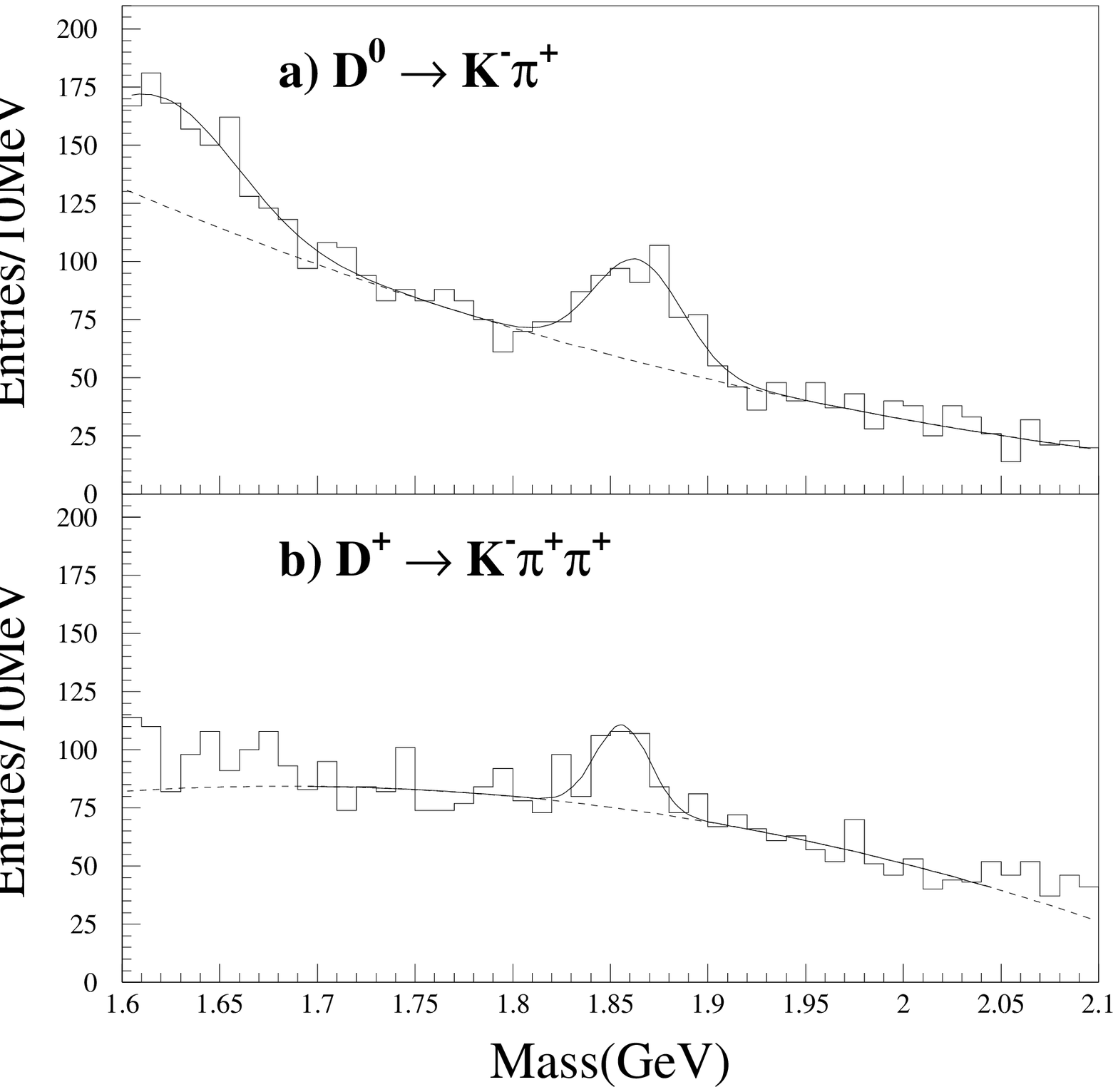,width=\textwidth,height=\textwidth}
\caption{The invariant mass of a) $K \pi$ and b) $K \pi \pi$ tags
with a momentum between 0 and 1 GeV at $E_{\rm c.m.} = 4.14$ GeV.}
\label{fig:2}
\end{besfig} 
\begin{besfig}
\besfigscale{0.1}
\psfig{file=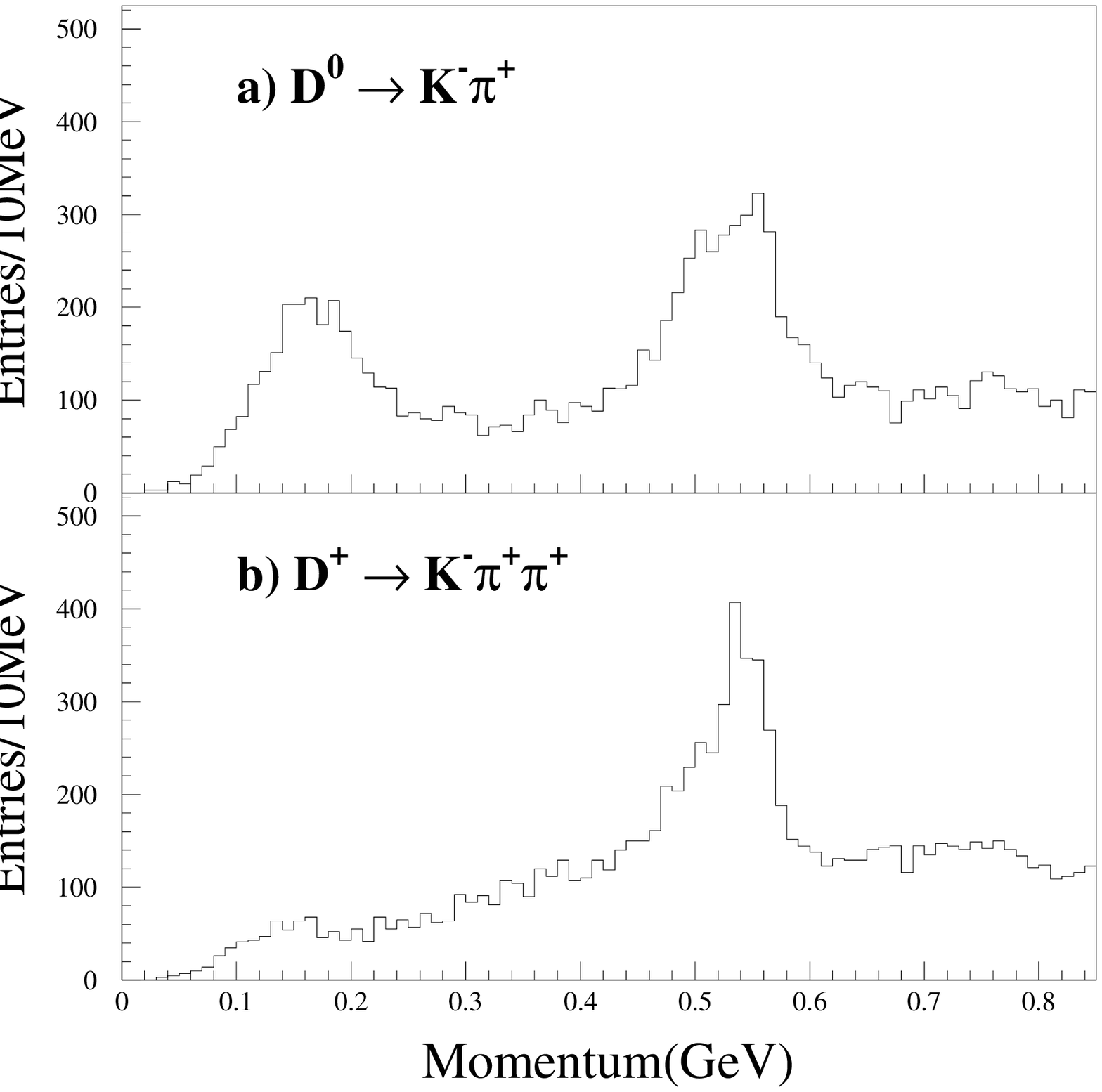,width=\textwidth,height=\textwidth}
\caption{The momentum of a) $K \pi$ and b) $K \pi \pi$ tags with a mass 
between 1.81 and 1.91 GeV at $E_{\rm c.m.} = 4.03$ GeV.}
\label{fig:3}
\end{besfig} 
\begin{besfig}
\besfigscale{0.5}
\psfig{file=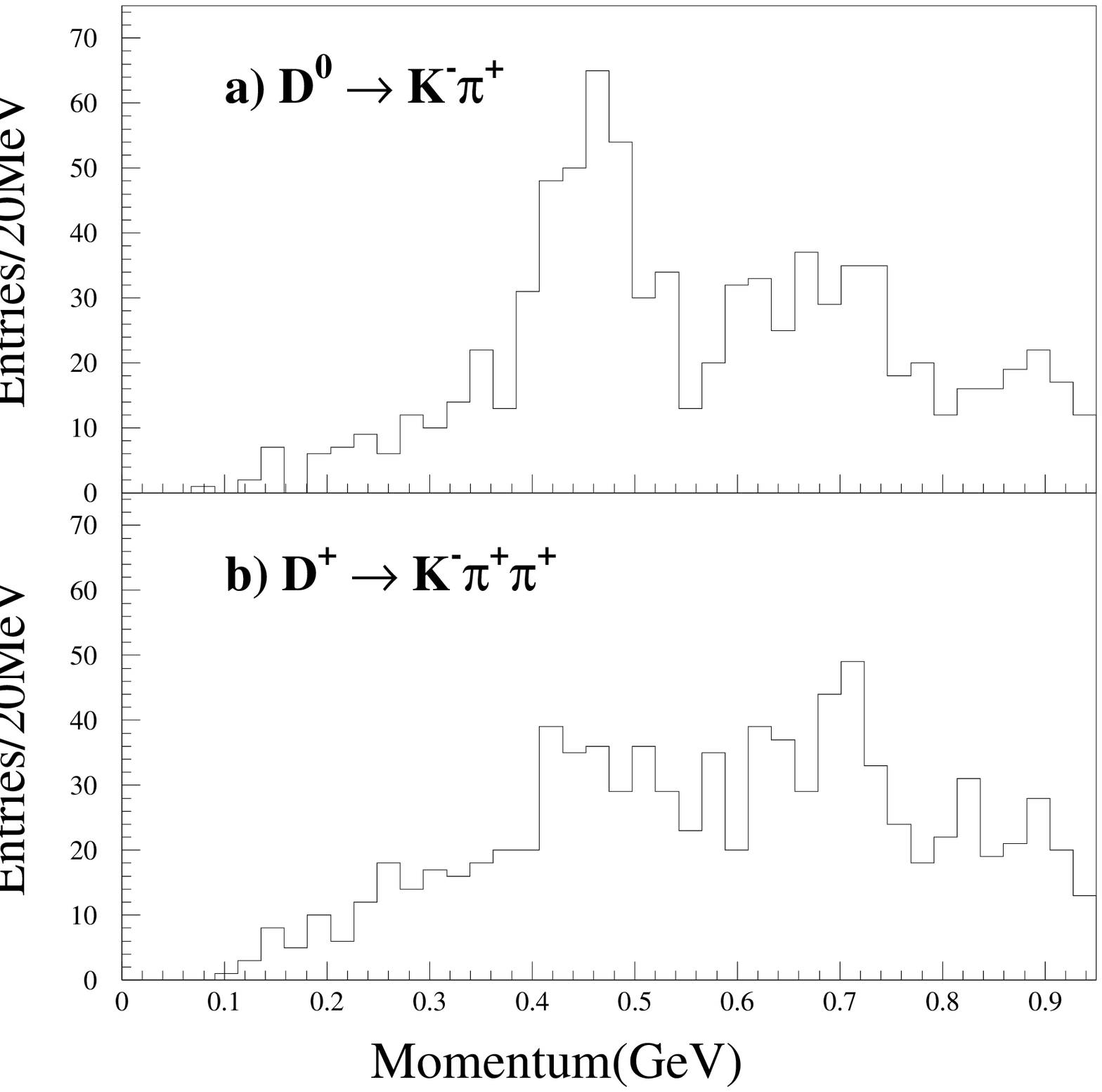,width=\textwidth,height=\textwidth}
\caption{The momentum of a) $K \pi$ and b) $K \pi \pi$ tags with a mass 
between 1.82 and 1.92 GeV at $E_{\rm c.m.} = 4.14$ GeV.}
\label{fig:4}
\end{besfig}  
\begin{besfig}
\besfigscale{0.5}
\psfig{file=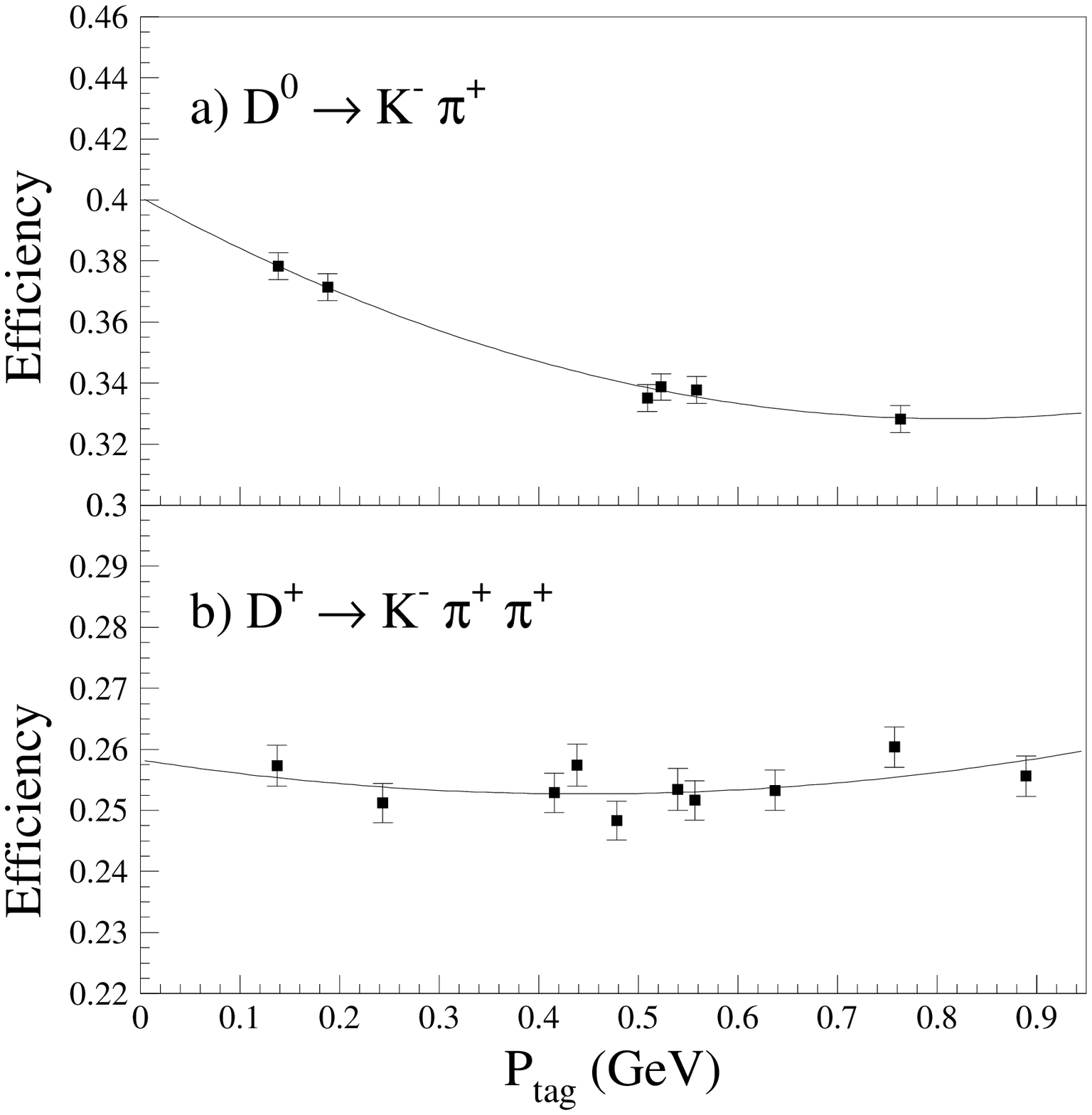,width=\textwidth,height=\textwidth}
\caption{The momentum dependence of the reconstruction efficiency.}
\label{fig:5}
\end{besfig} 
\begin{besfig}
\besfigscale{0.5}
\psfig{file=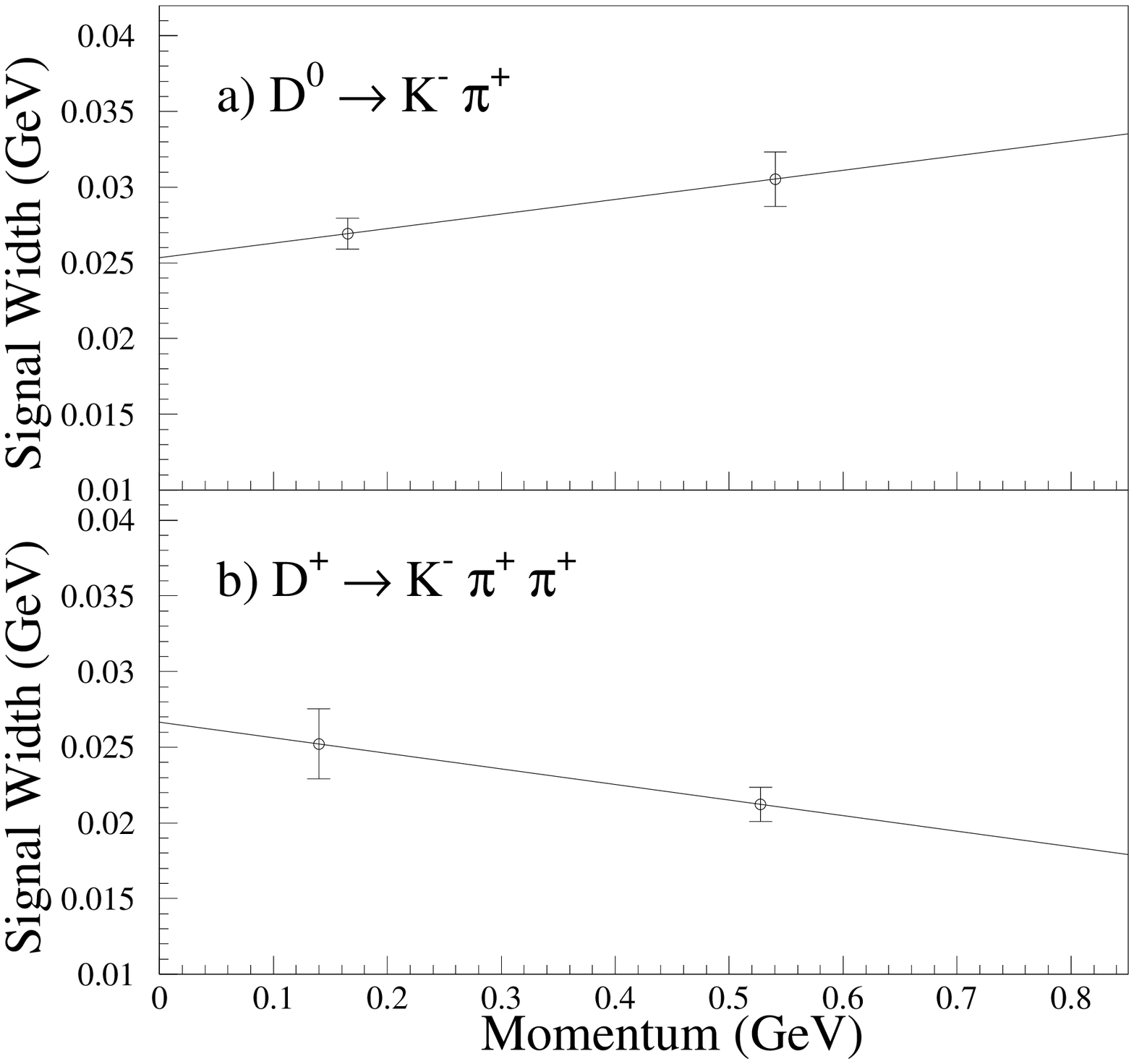,width=\textwidth,height=\textwidth}
\caption{The momentum dependence of the mass resolution.}
\label{fig:6}
\end{besfig} 
\begin{besfig}
\besfigscale{0.5}
\psfig{file=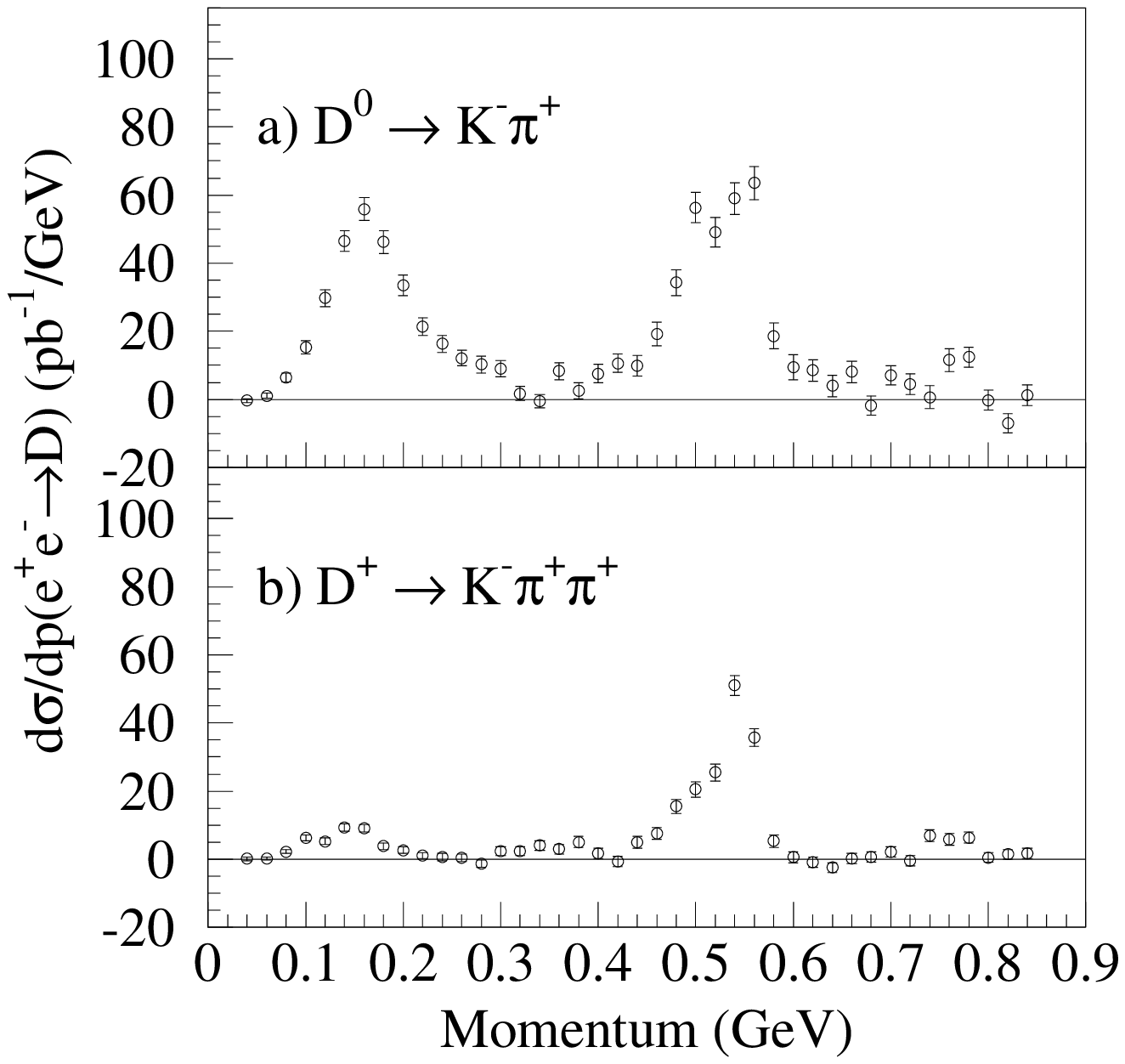,width=\textwidth,height=\textwidth}
\caption{
The differential production cross section of a) $e^+e^- \rightarrow D^0 X$ and b) $e^+e^- \rightarrow D^+ X$ versus the momentum of the reconstructed $D$ at $E_{\rm c.m.} = 4.03$ GeV.}
\label{fig:7}
\end{besfig}  
\begin{besfig}
\besfigscale{0.5}
\psfig{file=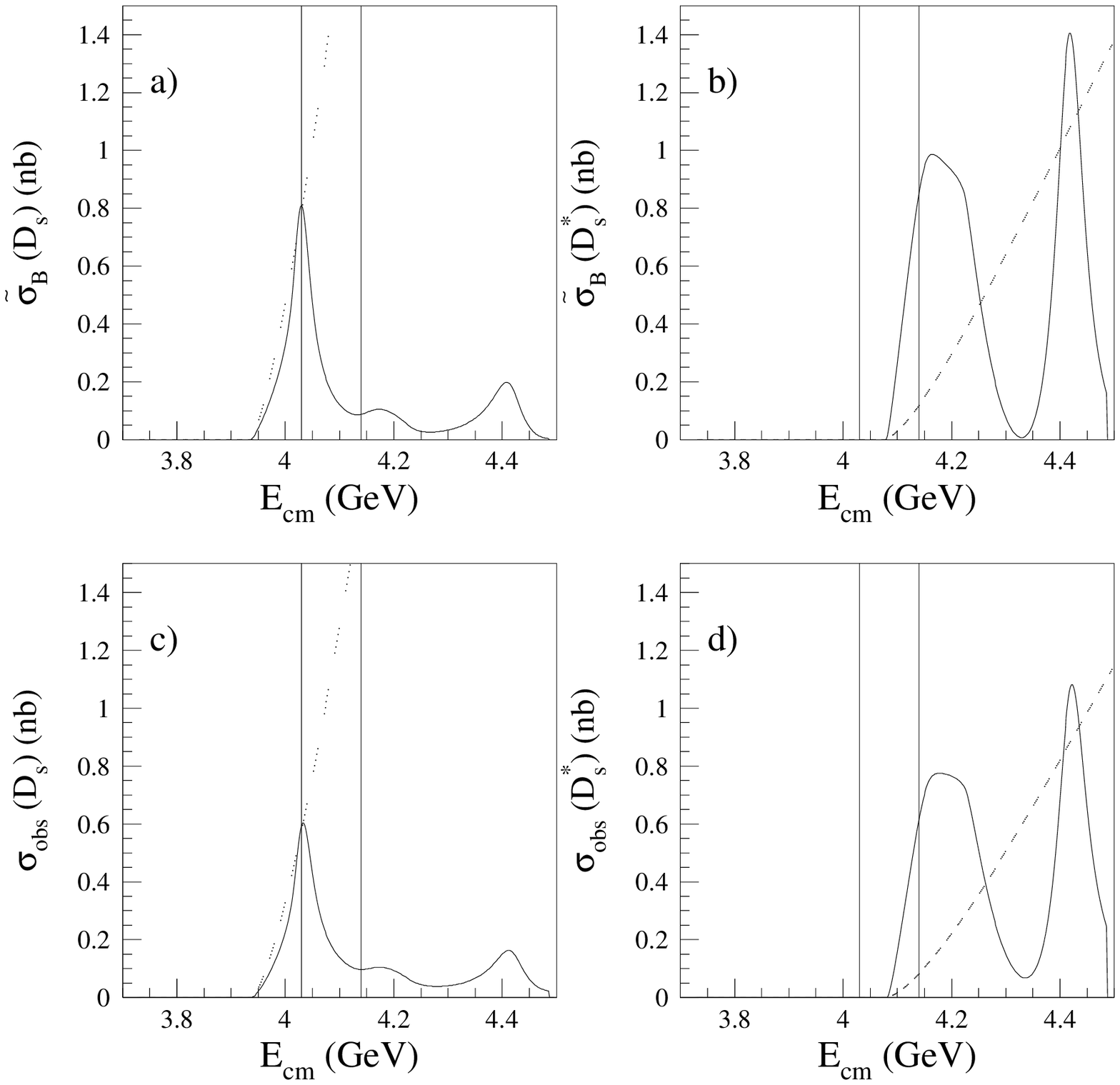,width=\textwidth,height=\textwidth}
\caption{
The differential production cross section of a) $e^+e^- \rightarrow D^0 X$ and b) $e^+e^- \rightarrow D^+ X$ versus the momentum of the reconstructed $D$ at $E_{\rm c.m.} = 4.14$ GeV.}
\label{fig:8}
\end{besfig} 

\begin{besfig}
\psfig{file=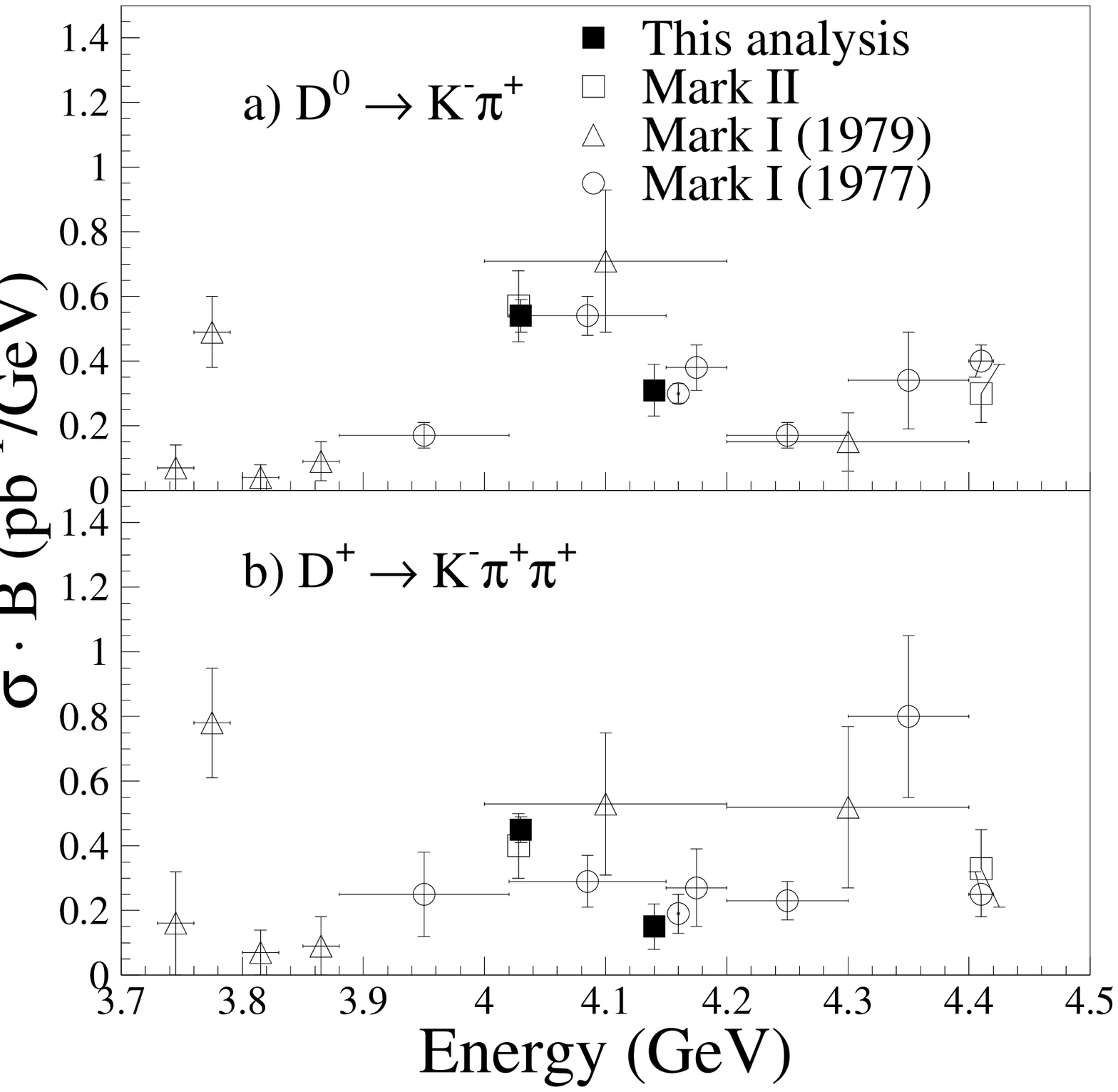,width=\columnwidth,height=\columnwidth}
\caption{
Comparison of the observed integrated cross section times branching
fraction for this analysis with those of Mark I and Mark II.
a) $D^0 \rightarrow K \pi$, b) $D^+ \rightarrow K\pi\pi$.}
\label{prev:1}
\end{besfig}

\begin{besfig}
\psfig{file=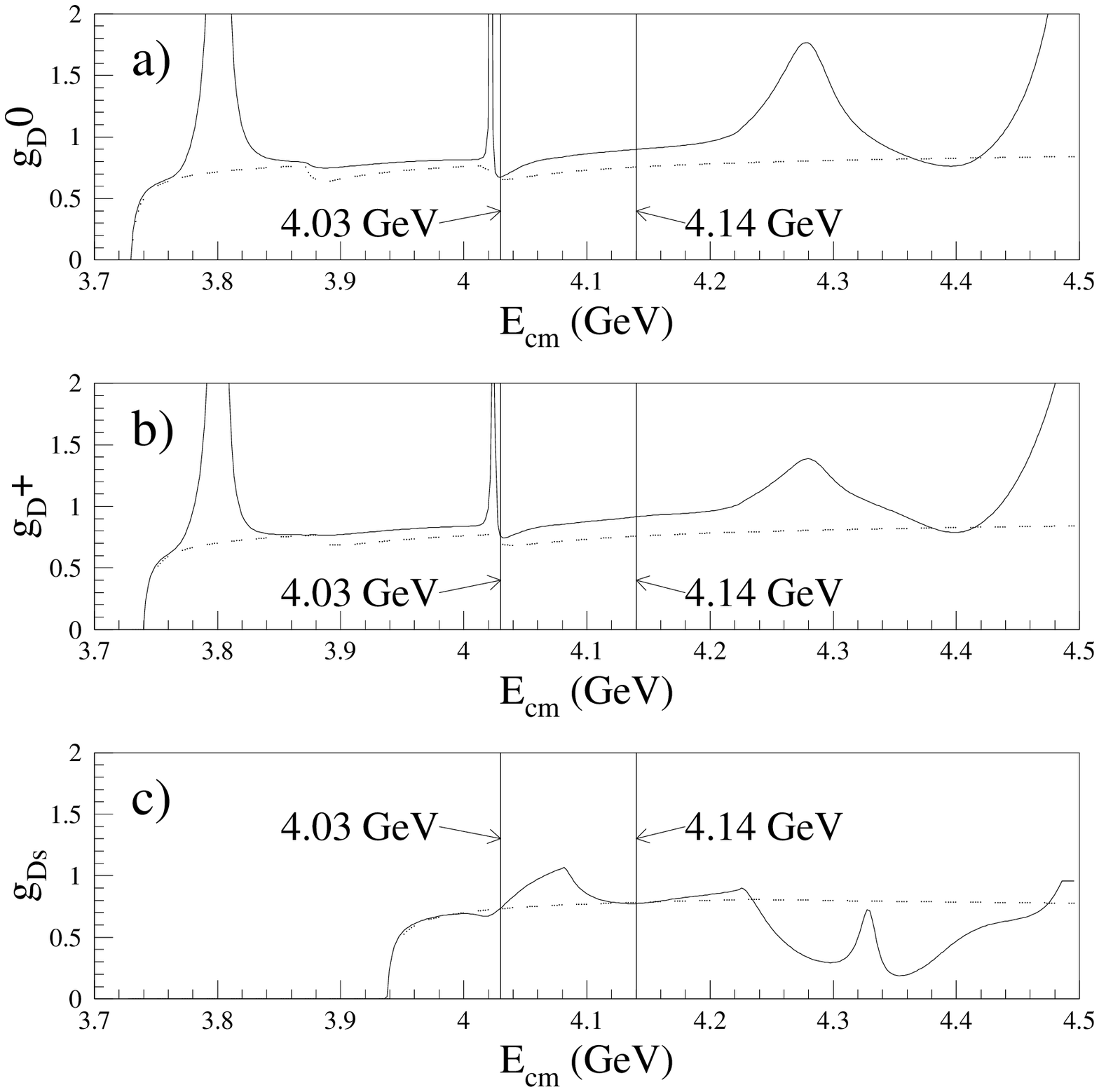,width=\textwidth,height=\textwidth}
\caption{The ISR correction for a) $D^0$, b) $D^+$, and c) $D_s$.  The 
correction using the Coupled Channel model is solid, while the
p-wave phase space formalism is dashed.
}
\label{isr:5}
\end{besfig}

\begin{besfig}
\psfig{file=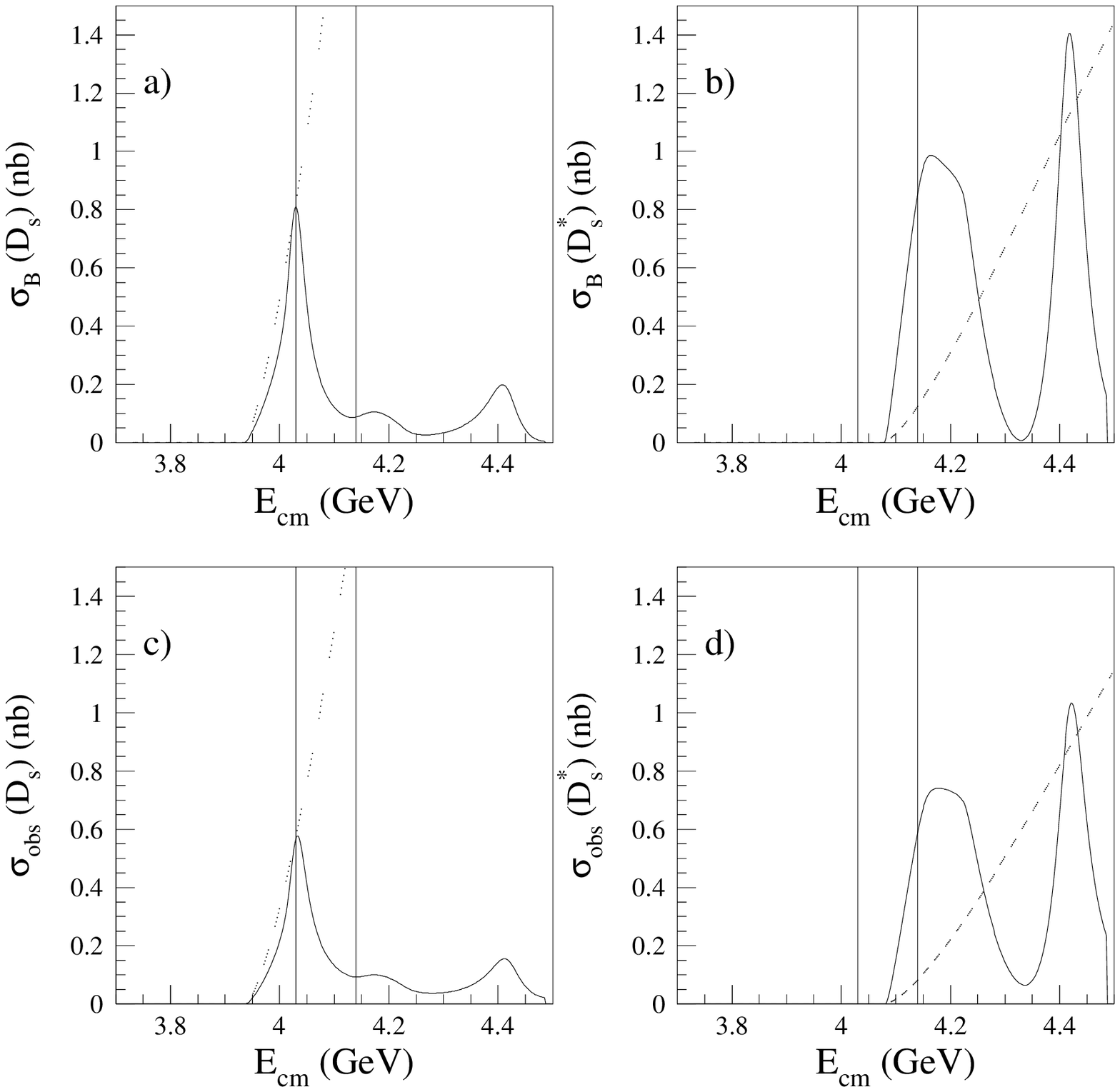,width=\columnwidth,height=\columnwidth
}
\caption{
a) The tree level cross section for $D_s$ mesons.  
b) The tree level cross section for $D_s^*$ mesons.  
c) The observed cross section for $D_s$ mesons.  
d) The observed cross section for $D_s^*$ mesons.  
The solid curve was obtained using the Coupled Channel model.  The dashed
line was obtained using the p-wave phase space formalism.
}
\label{isr:ds}
\end{besfig}

\begin{besfig}
\besfigscale{0.5}
\psfig{file=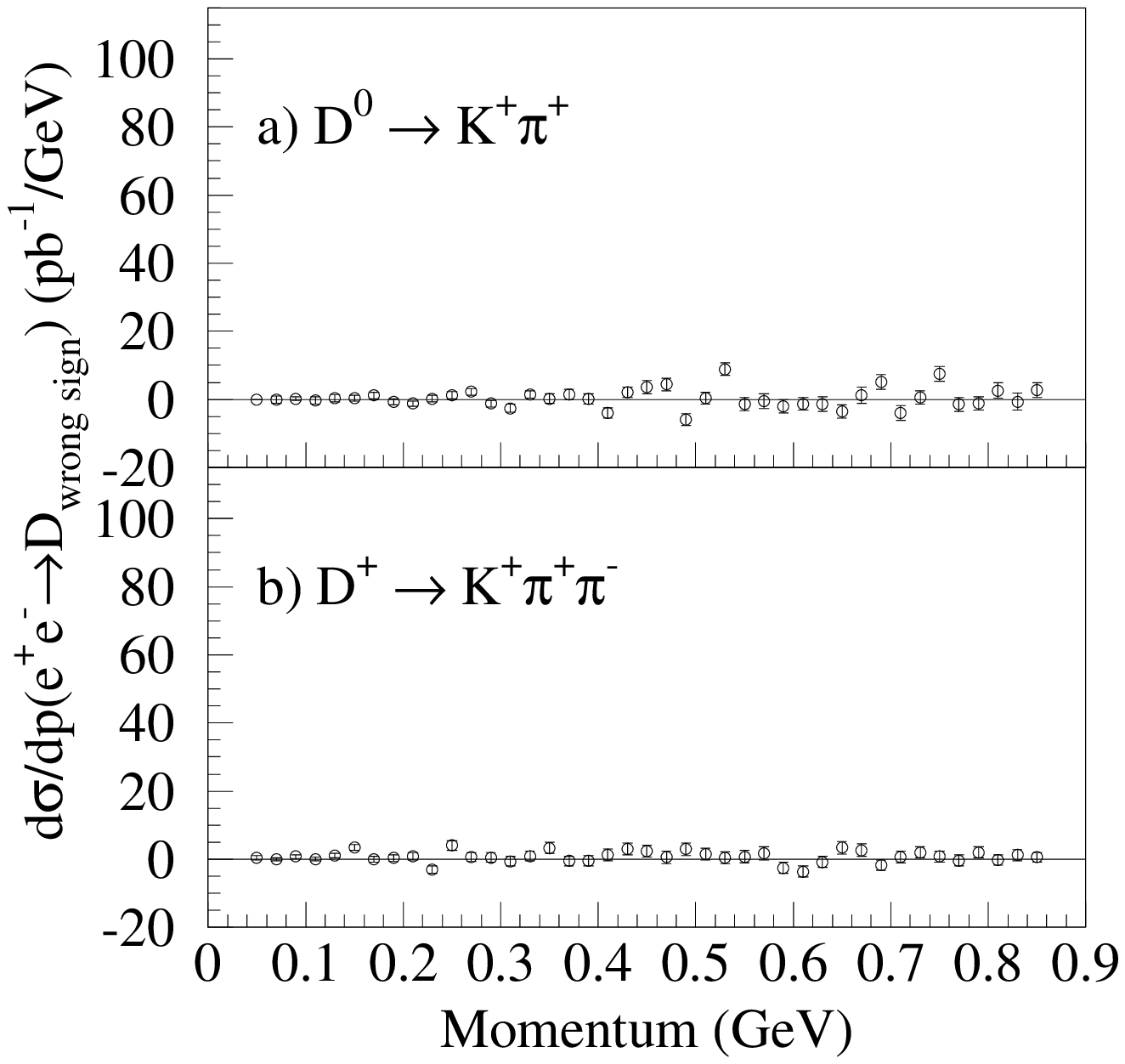,width=\textwidth,height=\textwidth}
\caption{
The differential production cross section for wrong sign
combinations.}
\label{fig:9}
\end{besfig} 

\begin{besfig}
\psfig{file=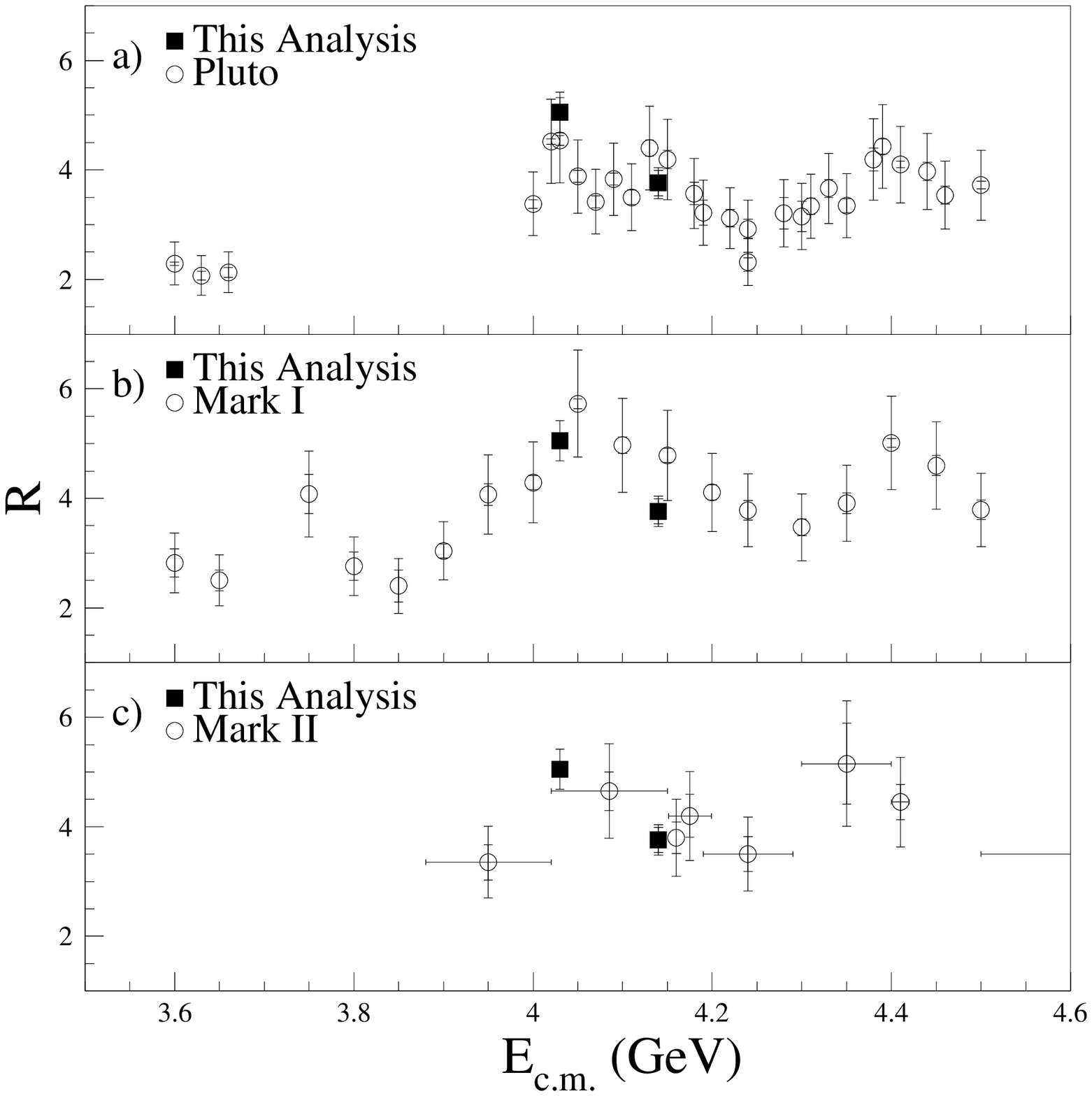,width=\columnwidth,height=\columnwidth}
\caption{
A comparison of the value of $R$ from this measurement with direct $R$ 
measurements from Pluto and Mark I and a charm-counting $R$ measurement
from Mark II.}
\label{prev:2}
\end{besfig}

\end{document}